\documentclass[nofootinbib,groupedaddress,superscriptaddress,unsortedaddress,showpacs]{article}
\usepackage{hyperref}
\usepackage{amsmath,amsfonts,amssymb}
\usepackage{graphicx}
\usepackage{epsfig,epsf}
\usepackage{bm}
\usepackage{float}
\restylefloat{table}

\def\opcit(#1){ {\em op. cit.}, #1}

\def\issue(#1,#2,#3){#1, #2 (#3)} 

\def\equationautorefname~#1\null{Eq.\,(#1)\null}
\def\pageautorefname\nobreakspace{p.}

\makeatletter\renewcommand{\p@subsection}{\thesection.}\makeatother%
 
\begin{document}

\renewcommand*{\thefootnote}{\fnsymbol{footnote}}


\begin{center}
{\Large\bf{Fermionic decay of light charged pseudoscalar in Georgi Machacek model}}


\vspace{5mm}

{\bf Swagata Ghosh}$^{a,b,c,}$\footnote{swgtghsh54@gmail.com}

\vspace{3mm}
{\em{${}^a$\ \ \ Department of Physics, Indian Institute of Technology Kharagpur, Kharagpur 721302, India.
}}

{\em{${}^b$\ \ \ Department of Physics and Astrophysics, University of 
Delhi, Delhi, India.
}}

{\em{${}^c$\ \ \ SGTB Khalsa College, University of Delhi, Delhi, India.}}

\end{center}

\begin{abstract}

The Georgi Machacek (GM) model is a triplet extension of the scalar sector of the Standard Model (SM) containing charged and neutral scalars and pseudoscalars with alluring phenomenology. 
The CMS and ATLAS collaborations of the LHC at $\sqrt{s}=13$ TeV already searched for various decays of the charged Higgs boson in low as well as high mass region. 
The low mass charged Higgs are produced by the decay of top quarks or antiquarks. 
$H^{\pm}\rightarrow \tau\nu,\, cs$ are the main fermionic decay channels of light charged scalars or pseudoscalars. 
For low mass region of the charged pseudoscalars, the triplet vev is restricted from above from the indirect search coming from the $b\rightarrow s \gamma$ decay. 
Here, I consider the observed data from the ATLAS and CMS both for the light charged Higgs decaying into $cs$ and $\tau\nu$. 
The ATLAS data observed for $H^{\pm}\rightarrow \tau\nu$ offers a more stringent bound on $v_2$ for GM model pseudoscalar mass below $160$ GeV. 

\end{abstract}

\setcounter{footnote}{0}
\renewcommand*{\thefootnote}{\arabic{footnote}}

\section{The Georgi Machacek Model}
\noindent
The Georgi-Machacek (GM) model \cite{Georgi:1985nv} extends the scalar sector of the Standard Model (SM) with one additional real triplet $\xi$ with hypercharge $0$ and one additional complex triplet $\chi$ with hypercharge $2$. 
These $SU(2)_L$ triplets can be expressed in terms of bi-triplet and the SM doublet can also be expressed as a bi-doublet, respectively, as :  
{\small
\begin{equation}
X= 
\begin{pmatrix}
 \chi^{0*} & \xi^{+} & \chi^{++}\cr
 \chi^{-} & \xi^0 & \chi^{+}\cr
 \chi^{--} & \xi^{-} & \chi^0
\end{pmatrix}\, ,
\;\;\qquad
\Phi= 
\begin{pmatrix}
\phi^{0*} & \phi^{+}\cr
\phi^{-} & \phi^0
\end{pmatrix}
.
\end{equation}
}
The equality of the vevs of the two triplets, $i.e.$ $v_{\xi} = v_{\chi} = v_2$, leads to the preservation of the custodial symmetry at tree level $(\rho_{tree}=1)$, such that, $v_1^2+8v_2^2 = v^2$, where, $v_1/\sqrt{2}$ and $v$ are the doublet vev and EW vev respectively. 
Also, $\tan \beta = 2\sqrt{2}v_2/v_1$ is considered. 

The scalar potential in terms of the bi-doublet and bi-triplet can generically be expressed as:
{\small
\begin{align}
\label{eq:pot}
\mathcal V \left(\Phi,X\right)
&=
\frac{{\mu_2}^2}{2}\, {\rm Tr}\left(\Phi^\dag\Phi\right)
+ \frac{{\mu_3}^2}{2}\, \rm{Tr}\left(X^\dag X\right)
+ {\lambda_1}\left[{\rm Tr}\left(\Phi^\dag\Phi\right)\right]^2
+ {\lambda_2}\, {\rm Tr}\left(\Phi^\dag\Phi\right)\, {\rm Tr}\left(X^\dag X\right)
\nonumber\\
&
+ {\lambda_3}\, {\rm Tr}\left(X^\dag X X^\dag X\right)
+ {\lambda_4}\left[{\rm Tr}\left(X^\dag X\right)\right]^2
- \frac{\lambda_5}{4}\, {\rm Tr}\left(\Phi^\dag \sigma^a \Phi\sigma^b\right)\, {\rm Tr}\left(X^\dag t^a X t^b\right)
\nonumber\\
&
- \frac{M_1}{4} \, {\rm Tr}\left(\Phi^\dag \sigma^a \Phi\sigma^b\right) {\left(U X U^\dag\right)_{ab}}
- {M_2} \, {\rm Tr}\left(X^\dag t^a X t^b\right) {\left(U X U^\dag\right)_{ab}}\, ,
\end{align}
}
with the Pauli matrices $\sigma^a$ and 
{\small
\begin{align}
t^1=\frac{1}{\sqrt2}
\begin{pmatrix}
 0 & 1 & 0\cr
 1 & 0 & 1\cr
 0 & 1 & 0
\end{pmatrix}
\,,\;\;
t^2=\frac{1}{\sqrt2}
\begin{pmatrix}
 0 & -i & 0\cr
 i & 0 & -i\cr
 0 & i & 0
\end{pmatrix}
\,,\;\;
t^3=
\begin{pmatrix}
 1 & 0 & 0\cr
 0 & 0 & 0\cr
 0 & 0 & -1
\end{pmatrix}\,,\;\;
U=
\frac{1}{\sqrt{2}}\begin{pmatrix}
 -1 & 0 & 1\cr
 -i & 0 & -i \cr
 0 & \sqrt{2} & 0
\end{pmatrix}\,.
\end{align}
}
The $SU(2)_L$ doublet and triplets give us the custodial fiveplet $(H_5^{\pm\pm,\,\pm,\,0})$, triplet $(H_3^{\pm,\,0})$, and two singlets $(h,H)$. $H_3^{\pm,\,0}$ are pseudoscalars, whereas the rest are scalars. 
All the physical scalars and pseudoscalars whether they are charged or neutral are listed below :
{\small
\begin{align}
&H_5^{\pm\pm}=\chi^{\pm\pm}\,, \;\;
H_5^{\pm}=\frac{\left(\chi^{\pm}-\xi^{\pm}\right)}{\sqrt{2}}\,, \;\;
H_5^0=\sqrt{\frac23}\xi^0-\sqrt{\frac13}\chi^{0R}\,, \;\;
\nonumber\\
&H_3^{\pm}=-\sin {\beta}\, \phi^{\pm}+\cos {\beta}\, \frac{\left(\chi^{\pm}+\xi^{\pm}\right)}{\sqrt{2}}\,, \;\; 
H_3^0=-\sin {\beta}\, \phi^{0I}+\cos {\beta}\, \chi^{0I}\,, \;\;
\nonumber\\
&h = \cos {\alpha}\,\, \phi^{0R}-\sin {\alpha}\,\, \sqrt{\frac13}\xi^0+\sqrt{\frac23}\chi^{0R}\,, \;\;
H = \sin {\alpha}\,\, \phi^{0R}+\cos {\alpha}\,\, \sqrt{\frac13}\xi^0+\sqrt{\frac23}\chi^{0R}\,.
\label{eq:fields}
\end{align}
}
The members of the fiveplet and triplet are mass degenerate with the common mass 
$m_5 = \sqrt{\frac{M_1}{4{v_2}}{v_1}^2+12{M_2}{v_2}+\frac32{\lambda_5}{v_1}^2+8{\lambda_3}{v_2}^2}$, 
and 
$m_3 = \sqrt{\left(\frac{M_1}{4{v_2}}+\frac{\lambda_5}{2}\right)v^2}$. 
The square of the masses of $h$ and $H$ are given by, 
${m_{h,H}}^2=\bigg({{\cal M}_{11}}^2+{{\cal M}_{22}}^2\mp\sqrt{\left({{\cal M}_{11}}^2-
{{\cal M}_{22}}^2\right)^2
+4\left({{\cal M}_{12}}^2\right)^2}\bigg)/2$, 
with the components of the mass-squared matrix between $h$ and $H$, given by, 
${\cal M}_{11}^2 = 8 {\lambda_1}{v_1}^2$, 
${\cal M}_{12}^2 = {\cal M}_{21}^2 = \frac{\sqrt{3}}{2} [ -M_1 + 4$ $(2\lambda_2 - \lambda_5) v_2] v_1$, 
${\cal M}_{22}^2 = \frac{M_1 v_1^2}{4 v_2} - 6M_2v_2 + 8(\lambda_3+3\lambda_4) v_2^2$. 
In this work, $h$ is SM-like with mass $m_h\approx 125$ GeV and $m_H > m_h$. 
The mixing angle between $h$ and $H$ is 
$\alpha = \tan^{-1}(\frac{2 {{\cal M}_{12}}^2}{{{\cal M}_{22}}^2-{{\cal M}_{11}}^2})/2$. 
The couplings of new Higgs bosons are given in \cite{Hartling:2014xma}. 
 \begin{figure}
  \begin{center}
   \includegraphics[width= 8.42cm]{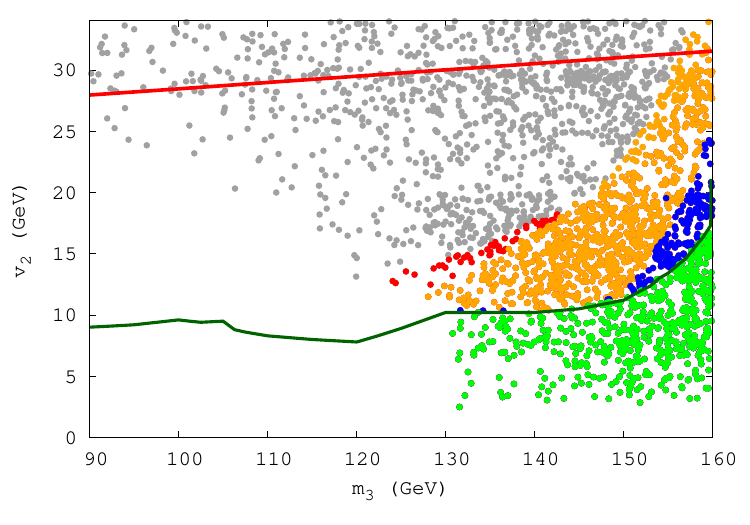} \\
  \end{center}
  \caption{\small The allowed parameter space in the $v_2-m_3$ plane.}
  \label{fig:m3v2}
 \end{figure}

\section{Constraints}
\noindent
The limits on the parameter space of the GM model come from theoretical constraints, LHC Higgs signal strengths, and the other indirect searches. 
The theoretical constraints including the perturbative unitarity and electroweak vacuum stability \cite{Hartling:2014zca} provide bounds on the quartic couplings $(\lambda_is)$. 
For the LHC signal strength at $\sqrt{s}=$13 TeV 
\cite{CMS:2018lkl,ATLAS:2019slw}, we consider $h$ as the SM-like Higgs. 
The contributions of the charged Higgs $H_3^{\pm}, H_5^{\pm\pm, \pm}$ in the $h \rightarrow \gamma \gamma$ decay constrain the parameter space of the GM model \cite{Ghosh:2022wbe}. 
All the points plotted in the Fig. \ref{fig:m3v2} are allowed by the above mentioned constraints coming from theory and LHC Higgs signal strength data.  
Before this work, the strongest limit on the triplet vev $v_2$ arose from the $b \rightarrow s \gamma$ observable \cite{Hartling:2014aga}. 
The red line in the Fig. \ref{fig:m3v2} indicates this limit. 

\section{Results}
\noindent
This paper considers the production of $H_3^{\pm}$ through the decay of $t(\overline{t})$ to $H_3^+b(H_3^-b)$ as only the decays of the charged Higgs with low masses are considered. 
The decay channels of light $H_3^{\pm}$ with two highest branching ratios are $H_3^{\pm} \rightarrow cs$ and $\tau^{\pm}\nu$ with $0.57$ and $0.39$ BR respectively. 
CMS \cite{CMS:2020osd} and ATLAS \cite{ATLAS:2024oqu} plotted the branching ratio (BR) of $t \rightarrow H^+b$ times the BR of $H^+ \rightarrow c\overline{s}$ as a function of the charged Higgs mass. 
CMS \cite{CMS:2019bfg} also searched for the production of $H^{\pm}$ times BR of $H^{\pm} \rightarrow \tau^{\pm} \nu_{\tau}$, whereas ATLAS \cite{ATLAS:2018gfm} searched for the BR of $t \rightarrow H^{\pm} b$ times the branching ratio of $H^{\pm} \rightarrow \tau^{\pm}\nu_{\tau}$, both as a function of the charged Higgs mass. 
All these searches were performed at $\sqrt{s}=13$ TeV. 
This paper uses the results of these experiments to provide the bound on triplet vev $v_2$ as the function of the mass $m_3$ of the charged Higgs $H_3^{\pm}$. 
The gray points under the red line in the Fig. \ref{fig:m3v2} are disallowed by these experimental results though they are allowed by the theoretical constraints, LHC Higgs signal strengths and $b \rightarrow s \gamma$ data as well as the results of the other indirect searches. 
All the colored points (red, orange, blue, green) are allowed by the CMS result corresponding to $H_3^{\pm} \rightarrow \tau^{\pm} \nu_{\tau}$. 
The red points are only allowed by this result but rejected by the other experimental results. 
The data from ATLAS experiments disrespect the orange points which are allowed by the CMS experimental results. 
The blue points are denied by the $H_3^{\pm} \rightarrow \tau^{\pm} \nu_{\tau}$ ATLAS data while they are permitted by the other results. 
Finally, the green points are allowed by all of the experimental data. 
The green solid line indicates the limit of the triplet vev $v_2$ as a function of the triplet mass $m_3$ corresponding to the ATLAS and CMS data for $H_3^{\pm} \rightarrow cs,  \tau^{\pm}\nu_{\tau}$. 
This green line sets the new limit on $v_2$, which is much stronger than the limit given by $b \rightarrow s \gamma$ data indicated by the red line in the Fig. \ref{fig:m3v2}. 

\section*{Acknowledgements}
\noindent
This work has been supported by Department of Science and Technology, Government of India through SERB-NPDF scholarship with grant no. PDF/2022/001784. 
The author would also like to acknowledge SERB grant CRG/2018/004889 as an honorary fellow without any financial support.

%
%

\end{document}